\documentclass[runningheads]{llncs}
\usepackage{graphicx}
\usepackage{hyperref}
\usepackage{xcolor}
\usepackage{xspace}
\usepackage{tikz}
\usepackage{float}
\usepackage{cleveref}

\begin{document}
\newcommand{\eg}{e.g.,\xspace}
\newcommand{\ie}{\textit{i.e.},\xspace}
\newcommand{\resp}{resp.,\xspace}
\newcommand{\vitamin}{\texttt{VITAMIN}\xspace}
\newcommand{\vadim}[1]{\textcolor{red}{Vadim: #1}\xspace}
\newcommand{\angelo}[1]{\textcolor{blue}{Angelo: #1}\xspace}
\title{\vitamin: A Compositional Framework for \\ Model Checking of Multi-Agent Systems}
%
\titlerunning{\vitamin: A Compositional Framework for Model Checking of MAS}
%
\author{Angelo Ferrando\inst{1}\orcidID{0000-0002-8711-4670} \and
Vadim Malvone\inst{2}\orcidID{0000-0001-6138-4229}}
\authorrunning{Angelo Ferrando \& Vadim Malvone}
%
\institute{
University of Modena and Reggio Emilia, Italy\\ \email{angelo.ferrando@unimore.it} \and Télécom Paris, France\\
\email{vadim.malvone@telecom-paris.fr}
}
\maketitle              
\begin{abstract}
The verification of Multi-Agent Systems (MAS) poses a significant challenge. Various approaches and methodologies exist to address this challenge; however, tools that support them are not always readily available. Even when such tools are accessible, they tend to be hard-coded, lacking in compositionality, and challenging to use due to a steep learning curve. In this paper, we introduce a methodology designed for the formal verification of MAS in a modular and versatile manner, along with an initial prototype, that we named \vitamin. Unlike existing verification methodologies and frameworks for MAS, \vitamin is constructed for easy extension to accommodate various logics (for specifying the properties to verify) and models (for determining on what to verify such properties).

\end{abstract}
\section{Introduction}

Software and hardware systems are notoriously challenging to verify. This difficulty generally arises from their complexity, making formalisation and proper analysis arduous. At times, it is due to their size, rendering exhaustive verification impractical unless appropriately abstracted or optimised (\eg through symbolic techniques). Regardless of the cause, formally verifying software and hardware systems is a complex task demanding deep expertise in formal methods. Given that such expertise is often scarce, formal verification techniques find limited usability in real-world software and hardware development.

Moving from monolithic systems to Multi-Agent Systems (MAS), formal verification becomes even more complex to achieve.
In fact, the process of testing~\cite{DBLP:conf/aose/NguyenPBPT09}, debugging~\cite{DBLP:conf/atal/Winikoff17}, and verifying~\cite{DBLP:journals/ase/DennisFWB12} such systems can be quite complex. Solutions which make
the development more reliable are of uttermost importance.
Similar to the challenges mentioned for monolithic systems, MASs encounter the same issues in verification. Moreover, as distributed systems comprising intelligent and independent components (the agents), their verification becomes even more demanding. This is due to the fact that MAS properties may rely on the rationality of the agents and on how they interact with each other.

One significant development in formal verification is \emph{Alternating-Time Temporal Logic} (ATL)~\cite{DBLP:journals/jacm/AlurHK02}, enabling reasoning about agents' strategies with temporal goals as payoff. However, ATL's implicit treatment of strategies limits its suitability for certain concepts, leading to the introduction of more powerful formalisms like \emph{Strategy Logic} (SL)~\cite{DBLP:journals/tocl/MogaveroMPV14}. SL treats strategies as first-order objects, providing a richer framework for strategic reasoning.
While SL's expressivity is high, it comes at the cost of non-elementary complete model-checking and undecidable satisfiability~\cite{DBLP:journals/tocl/MogaveroMPV14}. To address this, fragments like Strategy Logic with Simple-Goals~\cite{DBLP:conf/ijcai/BelardinelliJKM19} have been proposed, offering better computational properties while still subsuming ATL.
In the context of MAS, considering agents' visibility is crucial. The distinction between \emph{perfect} and \emph{imperfect} information MAS impacts model-checking complexity, with imperfect information scenarios often modelled using indistinguishability relations over MAS states~\cite{DBLP:journals/jcss/Reif84}. This distinction becomes particularly relevant, for instance, in rendering ATL undecidable in the context of imperfect information and memoryful strategies~\cite{DBLP:journals/corr/abs-1102-4225}. To overcome this problem, some works have either focused on an approximation to perfect information \cite{BelardinelliFM23}, developed notions of bounded memory \cite{BelardinelliLMY22}, or developed hybrid techniques \cite{FerrandoM22,FerrandoM23}.

Even with strong theoretical foundations, the formal verification of MAS heavily depends on tools that support such techniques. Notably, some tools stand out as pillars in this field, including MCMAS~\cite{DBLP:journals/sttt/LomuscioQR17} and STV~\cite{DBLP:conf/atal/KurpiewskiJK19}.

MCMAS is recognized as one of the most widely used model checkers for the strategic verification of multi-agent systems, primarily due to being one of the earliest tools developed, which served as a foundational proof-of-concept for researchers.
Despite the widespread use of MCMAS in the academic community, it exhibits issues that hinder its broader adoption, particularly outside the MAS research community itself. Specifically, its verification process is inherently hard-coded.
In fact, even though MCMAS has been extended in various ways, it lacks modularity and does not allow a clear separation between different logics and models that causes maintainability issues. 
That is, 
MCMAS lacks the capability of being transparently extended with new logics and models for the verification of MASs. 
%
Furthermore, while it does offer a graphical interface, users may find its execution challenging, as it requires additional tools, such as Eclipse, for installation. Moreover, the tool lacks comprehensive external documentation to assist developers, and its internal documentation may prove helpful only to those familiar with its original development. 
It is important to acknowledge that the observed limitations in MCMAS arise from its primary function as a research tool dedicated to proving theoretical contributions.
Regarding STV, it is designed to address specific verification goals in a predetermined way. Consequently, the resulting tool lacks compositional nature and only supports certain types of logics and models for the MAS verification. If users wish to verify the MAS against different logics or models, such flexibility is unavailable to them. Moreover, the tool lacks comprehensive documentation to assist users and developers.
Additionally, both MCMAS and STV require a strong background in formal methods, making them challenging for non-expert users to employ them. In summary, although widely used and with a history of successes, both MCMAS and STV tend to lack modularity and usability.

These two aspects are the ones we decided to tackle by engineering and developing a formal verification framework for MAS, called \vitamin (\texttt{V}er\texttt{I}fica\texttt{T}ion of \texttt{A} \texttt{M}ult\texttt{I}-age\texttt{N}t system), which aims at being both highly compositional, in terms of the logic and model formalisms that can be employed, and highly usable, in terms of the user experience (from a developer and end-user side). 
%
%
The concept behind our methodology is to generalise MAS verification without being tied to any specific logic or model formalism. \vitamin, even though still under development, achieves compositionality through its design, minimising assumptions about the types of logics and models that can be employed. Its end-user's usability is enhanced through a user experience that guides the entire verification process. It is worth noting that, \vitamin's compositionality facilitates straightforward extension of its components by external developers. 
In this paper, we refrain from presenting experimental results or benchmarks for two primary reasons. Firstly, an empirical evaluation would be beyond the scope of our work, which focuses on the engineering of \vitamin and its foundational aspects. Secondly, due to \vitamin's inherent compositionality, it can readily accommodate the integration of both new and existing verification techniques. Consequently, comparing \vitamin with existing verification tools for MAS would not be meaningful, as each tool would essentially be compared, in theory, with itself.

\section{\vitamin: architecture}


In this section, we focus on the engineering of our approach. To do so, we present an overview of the tool, named \vitamin, in~\Cref{fig:overview}, showcasing the main components of the latter. First and foremost, there is a clear separation between the logics and models used in our verification methodology. This separation of concerns is of paramount importance as it enables the tool to evolve independently in both directions. The logics and models should not be tied to each other, providing a more flexible environment where different logics can be verified on distinct models.

\tikzset{every picture/.style={line width=0.75pt}} 

\begin{figure*}
\centering
\scalebox{0.54}{
\begin{tikzpicture}[x=0.75pt,y=0.75pt,yscale=-1,xscale=1]

\draw  [fill={rgb, 255:red, 80; green, 227; blue, 194 }  ,fill opacity=0.1 ] (413.5,92.51) -- (801.5,92.51) -- (801.5,371) -- (413.5,371) -- cycle ;
\draw  [fill={rgb, 255:red, 208; green, 2; blue, 27 }  ,fill opacity=0.4 ] (425,211.4) .. controls (425,204) and (431,198) .. (438.4,198) -- (552.1,198) .. controls (559.5,198) and (565.5,204) .. (565.5,211.4) -- (565.5,251.6) .. controls (565.5,259) and (559.5,265) .. (552.1,265) -- (438.4,265) .. controls (431,265) and (425,259) .. (425,251.6) -- cycle ;
\draw  [fill={rgb, 255:red, 184; green, 233; blue, 134 }  ,fill opacity=0.4 ] (650,211.4) .. controls (650,204) and (656,198) .. (663.4,198) -- (777.1,198) .. controls (784.5,198) and (790.5,204) .. (790.5,211.4) -- (790.5,251.6) .. controls (790.5,259) and (784.5,265) .. (777.1,265) -- (663.4,265) .. controls (656,265) and (650,259) .. (650,251.6) -- cycle ;
\draw  [fill={rgb, 255:red, 74; green, 144; blue, 226 }  ,fill opacity=0.1 ] (514,304.4) .. controls (514,297) and (520,291) .. (527.4,291) -- (684.1,291) .. controls (691.5,291) and (697.5,297) .. (697.5,304.4) -- (697.5,344.6) .. controls (697.5,352) and (691.5,358) .. (684.1,358) -- (527.4,358) .. controls (520,358) and (514,352) .. (514,344.6) -- cycle ;
\draw   (589,71) -- (599.88,71) -- (599.88,47) -- (621.63,47) -- (621.63,71) -- (632.5,71) -- (610.75,87) -- cycle ;
\draw  [fill={rgb, 255:red, 80; green, 227; blue, 194 }  ,fill opacity=0.1 ] (820.75,223.25) -- (820.75,234.13) -- (844.75,234.13) -- (844.75,255.88) -- (820.75,255.88) -- (820.75,266.75) -- (804.75,245) -- cycle ;
\draw  [fill={rgb, 255:red, 74; green, 144; blue, 226 }  ,fill opacity=0.1 ] (625.5,393) -- (614.63,393) -- (614.63,417) -- (592.88,417) -- (592.88,393) -- (582,393) -- (603.75,377) -- cycle ;
\draw  [fill={rgb, 255:red, 144; green, 19; blue, 254 }  ,fill opacity=0.1 ] (504.5,118.4) .. controls (504.5,111) and (510.5,105) .. (517.9,105) -- (694.1,105) .. controls (701.5,105) and (707.5,111) .. (707.5,118.4) -- (707.5,158.6) .. controls (707.5,166) and (701.5,172) .. (694.1,172) -- (517.9,172) .. controls (510.5,172) and (504.5,166) .. (504.5,158.6) -- cycle ;
\draw  [fill={rgb, 255:red, 248; green, 231; blue, 28 }  ,fill opacity=0.4 ] (504.5,118.4) .. controls (504.5,111) and (510.5,105) .. (517.9,105) -- (603.85,105) .. controls (611.25,105) and (617.25,111) .. (617.25,118.4) -- (617.25,158.6) .. controls (617.25,166) and (611.25,172) .. (603.85,172) -- (517.9,172) .. controls (510.5,172) and (504.5,166) .. (504.5,158.6) -- cycle ;
\draw  [fill={rgb, 255:red, 80; green, 227; blue, 194 }  ,fill opacity=0.1 ] (392.75,266.75) -- (392.75,255.88) -- (368.75,255.88) -- (368.75,234.13) -- (392.75,234.13) -- (392.75,223.25) -- (408.75,245) -- cycle ;

\draw (495.25,231.5) node  [font=\Large] [align=left] {Model};
\draw (720.25,231.5) node  [font=\Large] [align=left] {Logic};
\draw (605.75,324.5) node  [font=\Large] [align=left] {\begin{minipage}[lt]{101.5pt}\setlength\topsep{0pt}
\begin{center}
Model Checker\\Interface
\end{center}

\end{minipage}};
\draw (611.25,24.5) node  [font=\Large] [align=left] {User};
\draw (910.25,243.5) node  [font=\Large] [align=left] {\begin{minipage}[lt]{72.94pt}\setlength\topsep{0pt}
\begin{center}
High-Level\\Developer
\end{center}

\end{minipage}};
\draw (605.25,454.5) node  [font=\Large] [align=left] {\begin{minipage}[lt]{69.68pt}\setlength\topsep{0pt}
\begin{center}
High and Low-Level\\Developers
\end{center}

\end{minipage}};
\draw (660.25,138.5) node  [font=\Large] [align=left] {\begin{minipage}[lt]{44.6pt}\setlength\topsep{0pt}
\begin{center}
{\small Expert }\\{\small GUI}
\end{center}

\end{minipage}};
\draw (564.38,138.5) node  [font=\Large] [align=left] {\begin{minipage}[lt]{73.25pt}\setlength\topsep{0pt}
\begin{center}
{\small Non-Expert }\\{\small GUI}
\end{center}

\end{minipage}};
\draw (304.25,244.5) node  [font=\Large] [align=left] {\begin{minipage}[lt]{72.94pt}\setlength\topsep{0pt}
\begin{center}
High-Level\\Developer
\end{center}

\end{minipage}};

\end{tikzpicture}

}
\caption{\vitamin's architecture.}
\label{fig:overview}
\end{figure*}

Another noteworthy aspect in~\Cref{fig:overview} is the presence of an interface dedicated to handling the actual formal verification. Further details on this aspect are discussed in the subsequent sections. It is crucial to emphasise that, similar to the logics and models, the verification component foreseen in our methodology is independent and highly compositional. As an illustration, we can choose to verify ATL properties on a Concurrent Game Structure (CGS)~\cite{DBLP:journals/jacm/AlurHK02}, and such verification can be executed in different ways. For instance, we may opt for an \texttt{explicit} verification based on fix-point (similar to what is done in the case of CTL properties), or alternatively, symbolically encode the model as a Binary Decision Diagram (BDD) and perform symbolic (\ie \texttt{implicit}) model checking, or finally, \texttt{abstract} the model to cope with its complexity and perform the verification on the resulting abstracted model. These examples underscore how the actual verification, considering a logic and a model, can be achieved in various ways. Importantly, this aspect is kept separate from the tool's ecosystem to avoid hard-coding within it.

One additional aspect to note in~\Cref{fig:overview} is the type of users envisioned in our approach. Rather than the standard end-user, we assume the presence of three user categories. The first one corresponds to what we commonly refer to as an end-user: a user who utilises the verification approach solely for verifying a MAS, without the intention or objective of extending or modifying the tool itself.
In addition to the standard \texttt{User}, we envisage the access to two levels of developer users: the \texttt{High-Level Developer}, who concentrates on formal verification aspects, and the \texttt{Low-Level Developer}, who focuses on optimisation and low-level implementation.

\texttt{High-Level Developer} users possess experience in model checking within MAS and are responsible for developing the \texttt{Model} and \texttt{Logic} components and can develop the explicit verification via \texttt{Model Checker Interface}. They can achieve this by extending existing models and logics or by introducing entirely new ones into the architecture. Importantly, as each model in the \texttt{Model} component, each logic in the \texttt{Logic} component, and each explicit verification algorithm in the \texttt{Model Checker Interface} component is developed as an independent module, the enrichment of these components with new modules does not introduce errors into previously validated modules. These developers are concerned with properly defining and verifying the models and logics in the system, without delving into the intricacies of creating high-performance software solutions.

The task of optimising such implementations is undertaken by \texttt{Low-Level Developer} users who possess expertise not only in model checking within MAS but also in software engineering. They are responsible for enhancing the implementation provided by \texttt{High-Level Developers} by leveraging optimisation techniques, which can encompass both algorithmic and data structure improvements.
For example, a \texttt{High-Level Developer} may propose a logic and its verification on an explicit model, and a \texttt{Low-Level Developer}, starting from such an implementation, may offer an optimised solution based on the verification of an implicit model, such as through BDDs.

To bridge the gap between these two types of developers, we have introduced the \texttt{Model Checker Interface} component in our methodology. This component, developed by low-level developers, is intended to be utilised by end-users and high-level developers to seamlessly use the optimisations.

Thanks to the presented architecture, the strengths of our approach include:
\begin{itemize}
    \item[\textbf{Modularity:}] it allows transparent extension without the need for core engine modifications.
    \item[\textbf{User-Friendlyness:}] it enables end-users to use the software without requiring expertise in formal verification.
    \item[\textbf{Distribution:}] the compositionality of the verification components enables their distribution on different machines.
    \item[\textbf{Documentation:}] we prioritise the development of internal and external documentation to assist users and developers in effectively utilising and extending the tool.
\end{itemize}

\section{\vitamin: detail on the architecture modules}


We now delve more into the details of the components of the architecture behind our verification methodology. To help us present it, we report a diagram for each of such components. Note that, the colours are consistent with the ones used in~\Cref{fig:overview}, which denotes the more general diagram of the architecture and its components in a whole.
Specifically, we can see how the compositionality is achieved through hierarchical structures; where in the root we have the general notion, and going deeper into the resulting hierarchy, we have various instantiations to serve different verification purposes.


\subsection{Model}


As illustrated in~\Cref{fig:model}, the model component is structured hierarchically. At the root, we find the concept of a model, while further down in the hierarchy, two crucial branches of system models emerge: \texttt{Monolithic} and \texttt{Multi-Agent}. The former represents the standard model for software and hardware systems, such as Kripke Structures~\cite{DBLP:books/cu/Chellas80} and Labelled Transition Systems~\cite{DBLP:journals/cacm/Keller76}. These models are commonly used to depict the behaviour of centralised systems. However, such models typically do not account for the presence of autonomous entities (such as agents) and lack a proper means to characterise their independent and rational behaviours. To address this aspect, the methodology also supports the notion of multi-agent models, including Concurrent Game Structures (CGSs)~\cite{DBLP:journals/jacm/AlurHK02} and Interpreted Systems (ISs)~\cite{DBLP:books/mit/FHMV1995}.

\tikzset{every picture/.style={line width=0.75pt}} 

\begin{figure*}
\centering
\scalebox{0.35}{
\begin{tikzpicture}[x=0.75pt,y=0.75pt,yscale=-1,xscale=1]

\draw  [fill={rgb, 255:red, 208; green, 2; blue, 27 }  ,fill opacity=0.4 ] (259.5,33.3) -- (815.5,33.3) -- (815.5,395.3) -- (259.5,395.3) -- cycle ;
\draw  [fill={rgb, 255:red, 255; green, 255; blue, 255 }  ,fill opacity=1 ] (453.59,50.7) .. controls (453.61,43.3) and (459.62,37.31) .. (467.02,37.33) -- (623.72,37.72) .. controls (631.12,37.73) and (637.11,43.75) .. (637.09,51.15) -- (636.99,91.35) .. controls (636.97,98.75) and (630.96,104.73) .. (623.56,104.72) -- (466.86,104.33) .. controls (459.46,104.31) and (453.47,98.3) .. (453.49,90.9) -- cycle ;
\draw  [fill={rgb, 255:red, 255; green, 255; blue, 255 }  ,fill opacity=1 ] (508.36,139.44) -- (518.9,125.47) -- (518.87,134.22) -- (538.12,134.26) -- (538.15,124.01) -- (529.4,123.99) -- (543.42,113.53) -- (557.4,124.06) -- (548.65,124.04) -- (548.62,134.29) -- (567.87,134.34) -- (567.9,125.59) -- (578.36,139.61) -- (567.83,153.59) -- (567.85,144.84) -- (518.85,144.72) -- (518.83,153.47) -- cycle ;
\draw  [fill={rgb, 255:red, 255; green, 255; blue, 255 }  ,fill opacity=1 ] (391.29,174.35) .. controls (391.3,166.39) and (397.77,159.96) .. (405.72,159.98) -- (526.42,160.28) .. controls (534.37,160.3) and (540.8,166.76) .. (540.78,174.71) -- (540.68,217.91) .. controls (540.66,225.87) and (534.2,232.3) .. (526.24,232.28) -- (405.54,231.98) .. controls (397.59,231.96) and (391.16,225.5) .. (391.18,217.55) -- cycle ;
\draw  [fill={rgb, 255:red, 255; green, 255; blue, 255 }  ,fill opacity=1 ] (550.28,174.74) .. controls (550.3,166.78) and (556.76,160.35) .. (564.72,160.37) -- (685.42,160.67) .. controls (693.37,160.69) and (699.8,167.15) .. (699.78,175.1) -- (699.68,218.3) .. controls (699.66,226.25) and (693.19,232.69) .. (685.24,232.67) -- (564.54,232.37) .. controls (556.59,232.35) and (550.16,225.89) .. (550.18,217.94) -- cycle ;
\draw  [fill={rgb, 255:red, 255; green, 255; blue, 255 }  ,fill opacity=1 ] (263.26,285.82) .. controls (263.28,279.74) and (268.22,274.82) .. (274.29,274.84) -- (381.55,275.1) .. controls (387.63,275.11) and (392.55,280.05) .. (392.53,286.13) -- (392.45,319.15) .. controls (392.44,325.23) and (387.5,330.14) .. (381.42,330.13) -- (274.16,329.87) .. controls (268.08,329.85) and (263.17,324.91) .. (263.18,318.83) -- cycle ;
\draw  [fill={rgb, 255:red, 255; green, 255; blue, 255 }  ,fill opacity=1 ] (401.26,286.82) .. controls (401.28,280.74) and (406.22,275.82) .. (412.29,275.84) -- (519.55,276.1) .. controls (525.63,276.11) and (530.55,281.05) .. (530.53,287.13) -- (530.45,320.15) .. controls (530.44,326.23) and (525.5,331.14) .. (519.42,331.13) -- (412.16,330.87) .. controls (406.08,330.85) and (401.17,325.91) .. (401.18,319.83) -- cycle ;
\draw  [fill={rgb, 255:red, 255; green, 255; blue, 255 }  ,fill opacity=1 ] (547.26,286.82) .. controls (547.28,280.74) and (552.22,275.82) .. (558.29,275.84) -- (665.55,276.1) .. controls (671.63,276.11) and (676.55,281.05) .. (676.53,287.13) -- (676.45,320.15) .. controls (676.44,326.23) and (671.5,331.14) .. (665.42,331.13) -- (558.16,330.87) .. controls (552.08,330.85) and (547.17,325.91) .. (547.18,319.83) -- cycle ;
\draw  [fill={rgb, 255:red, 255; green, 255; blue, 255 }  ,fill opacity=1 ] (683.26,286.82) .. controls (683.28,280.74) and (688.22,275.82) .. (694.29,275.84) -- (801.55,276.1) .. controls (807.63,276.11) and (812.55,281.05) .. (812.53,287.13) -- (812.45,320.15) .. controls (812.44,326.23) and (807.5,331.14) .. (801.42,331.13) -- (694.16,330.87) .. controls (688.08,330.85) and (683.17,325.91) .. (683.18,319.83) -- cycle ;
\draw  [fill={rgb, 255:red, 255; green, 255; blue, 255 }  ,fill opacity=1 ] (365.26,259.38) -- (375.79,245.41) -- (375.77,254.16) -- (395.02,254.21) -- (395.04,243.96) -- (386.29,243.94) -- (400.32,233.47) -- (414.29,244) -- (405.54,243.98) -- (405.52,254.23) -- (424.77,254.28) -- (424.79,245.53) -- (435.26,259.56) -- (424.72,273.53) -- (424.74,264.78) -- (375.74,264.66) -- (375.72,273.41) -- cycle ;
\draw  [fill={rgb, 255:red, 255; green, 255; blue, 255 }  ,fill opacity=1 ] (645.18,259.58) -- (655.71,245.61) -- (655.69,254.36) -- (674.94,254.4) -- (674.97,244.15) -- (666.22,244.13) -- (680.24,233.67) -- (694.22,244.2) -- (685.47,244.18) -- (685.44,254.43) -- (704.69,254.48) -- (704.71,245.73) -- (715.18,259.75) -- (704.64,273.73) -- (704.66,264.98) -- (655.66,264.86) -- (655.64,273.61) -- cycle ;
\draw  [fill={rgb, 255:red, 255; green, 255; blue, 255 }  ,fill opacity=1 ] (617.26,345.82) .. controls (617.28,339.74) and (622.22,334.82) .. (628.29,334.84) -- (735.55,335.1) .. controls (741.63,335.11) and (746.55,340.05) .. (746.53,346.13) -- (746.45,379.15) .. controls (746.44,385.23) and (741.5,390.14) .. (735.42,390.13) -- (628.16,389.87) .. controls (622.08,389.85) and (617.17,384.91) .. (617.18,378.83) -- cycle ;
\draw  [fill={rgb, 255:red, 255; green, 255; blue, 255 }  ,fill opacity=1 ] (334.26,346.82) .. controls (334.28,340.74) and (339.22,335.82) .. (345.29,335.84) -- (452.55,336.1) .. controls (458.63,336.11) and (463.55,341.05) .. (463.53,347.13) -- (463.45,380.15) .. controls (463.44,386.23) and (458.5,391.14) .. (452.42,391.13) -- (345.16,390.87) .. controls (339.08,390.85) and (334.17,385.91) .. (334.18,379.83) -- cycle ;

\draw (545.29,71.02) node  [font=\Large,rotate=-0.14] [align=left] {\begin{minipage}[lt]{42.73pt}\setlength\topsep{0pt}
\begin{center}
Model
\end{center}

\end{minipage}};
\draw (465.98,196.13) node  [font=\large,rotate=-0.14] [align=left] {Monolithic};
\draw (624.98,196.52) node  [font=\large,rotate=-0.14] [align=left] {Multi-Agent};
\draw (327.86,302.48) node  [font=\normalsize,rotate=-0.14] [align=left] {\begin{minipage}[lt]{44.12pt}\setlength\topsep{0pt}
\begin{center}
Kripke \\Structure
\end{center}

\end{minipage}};
\draw (465.86,303.48) node  [font=\normalsize,rotate=-0.14] [align=left] {\begin{minipage}[lt]{83.99pt}\setlength\topsep{0pt}
\begin{center}
Labelled \\Transition System
\end{center}

\end{minipage}};
\draw (611.86,303.48) node  [font=\normalsize,rotate=-0.14] [align=left] {\begin{minipage}[lt]{52.09pt}\setlength\topsep{0pt}
\begin{center}
Interpreted\\System
\end{center}

\end{minipage}};
\draw (747.86,305.48) node  [font=\normalsize,rotate=-0.14] [align=left] {\begin{minipage}[lt]{83.83pt}\setlength\topsep{0pt}
\begin{center}
Concurrent Game\\Structure
\end{center}

\end{minipage}};
\draw (681.86,362.48) node  [font=\normalsize,rotate=-0.14] [align=left] {\begin{minipage}[lt]{11.22pt}\setlength\topsep{0pt}
\begin{center}
...
\end{center}

\end{minipage}};
\draw (398.86,363.48) node  [font=\normalsize,rotate=-0.14] [align=left] {\begin{minipage}[lt]{11.22pt}\setlength\topsep{0pt}
\begin{center}
...
\end{center}

\end{minipage}};

\end{tikzpicture}

}
\caption{Model component insights.}
\label{fig:model}
\end{figure*}

It is worth noting that, we envision the possibility of further extending the structure of the hierarchy for the model component. Nevertheless, in its initial phase, we have chosen to consider the two most influential and commonly used branches of models, particularly for specifying system behaviour (at least for verification purposes).

\subsection{Logics}


As depicted in~\Cref{fig:logic}, an independent structure is in place to manage the various logics available. The necessity for such a hierarchy arises from the diverse nature of the logics that can be employed for the verification of Multi-Agent Systems (MAS). Specifically, in our methodology we distinguish between two types of formalisms for denoting properties to be verified: \texttt{Temporal} and \texttt{Strategic}. The former encompasses more standard temporal logics, such as Linear Temporal Logic (LTL)~\cite{DBLP:conf/focs/Pnueli77} or Computation Tree Logic (CTL)~\cite{DBLP:conf/lop/ClarkeE81}. The latter involves more recent strategic logics, with examples including Alternating-Time Temporal Logic (ATL)~\cite{DBLP:journals/jacm/AlurHK02} or Strategy Logic (SL)~\cite{DBLP:journals/tocl/MogaveroMPV14}.

\tikzset{every picture/.style={line width=0.75pt}} 

\begin{figure*}
\centering
\scalebox{0.35}{
\begin{tikzpicture}[x=0.75pt,y=0.75pt,yscale=-1,xscale=1]

\draw  [fill={rgb, 255:red, 184; green, 233; blue, 134 }  ,fill opacity=0.4 ] (260.5,18.3) -- (816.5,18.3) -- (816.5,380.3) -- (260.5,380.3) -- cycle ;
\draw  [fill={rgb, 255:red, 255; green, 255; blue, 255 }  ,fill opacity=1 ] (435.42,36.81) .. controls (435.37,29.41) and (441.34,23.38) .. (448.74,23.33) -- (605.44,22.43) .. controls (612.84,22.39) and (618.87,28.35) .. (618.91,35.75) -- (619.15,75.95) .. controls (619.19,83.35) and (613.22,89.39) .. (605.82,89.43) -- (449.13,90.33) .. controls (441.72,90.37) and (435.69,84.41) .. (435.65,77.01) -- cycle ;
\draw  [fill={rgb, 255:red, 255; green, 255; blue, 255 }  ,fill opacity=1 ] (494.95,127.07) -- (505.37,113.01) -- (505.42,121.76) -- (524.67,121.65) -- (524.61,111.4) -- (515.86,111.45) -- (529.8,100.87) -- (543.86,111.29) -- (535.11,111.34) -- (535.17,121.59) -- (554.42,121.48) -- (554.37,112.73) -- (564.95,126.67) -- (554.53,140.73) -- (554.48,131.98) -- (505.48,132.26) -- (505.53,141.01) -- cycle ;
\draw  [fill={rgb, 255:red, 255; green, 255; blue, 255 }  ,fill opacity=1 ] (379.13,159.64) .. controls (379.09,151.68) and (385.5,145.2) .. (393.45,145.15) -- (514.15,144.46) .. controls (522.1,144.41) and (528.58,150.82) .. (528.63,158.78) -- (528.88,201.98) .. controls (528.93,209.93) and (522.52,216.41) .. (514.56,216.46) -- (393.86,217.15) .. controls (385.91,217.2) and (379.43,210.79) .. (379.38,202.84) -- cycle ;
\draw  [fill={rgb, 255:red, 255; green, 255; blue, 255 }  ,fill opacity=1 ] (538.14,159.72) .. controls (538.09,151.77) and (544.5,145.28) .. (552.45,145.24) -- (673.15,144.54) .. controls (681.1,144.5) and (687.59,150.91) .. (687.63,158.86) -- (687.88,202.06) .. controls (687.93,210.01) and (681.52,216.5) .. (673.57,216.54) -- (552.87,217.24) .. controls (544.92,217.28) and (538.43,210.87) .. (538.39,202.92) -- cycle ;
\draw  [fill={rgb, 255:red, 255; green, 255; blue, 255 }  ,fill opacity=1 ] (264.26,272.02) .. controls (264.28,265.94) and (269.22,261.03) .. (275.29,261.04) -- (382.55,261.31) .. controls (388.63,261.32) and (393.55,266.26) .. (393.53,272.34) -- (393.45,305.36) .. controls (393.44,311.43) and (388.5,316.35) .. (382.42,316.33) -- (275.16,316.07) .. controls (269.08,316.06) and (264.17,311.12) .. (264.18,305.04) -- cycle ;
\draw  [fill={rgb, 255:red, 255; green, 255; blue, 255 }  ,fill opacity=1 ] (402.26,273.02) .. controls (402.28,266.94) and (407.22,262.03) .. (413.29,262.04) -- (520.55,262.31) .. controls (526.63,262.32) and (531.55,267.26) .. (531.53,273.34) -- (531.45,306.36) .. controls (531.44,312.43) and (526.5,317.35) .. (520.42,317.33) -- (413.16,317.07) .. controls (407.08,317.06) and (402.17,312.12) .. (402.18,306.04) -- cycle ;
\draw  [fill={rgb, 255:red, 255; green, 255; blue, 255 }  ,fill opacity=1 ] (548.26,273.02) .. controls (548.28,266.94) and (553.22,262.03) .. (559.29,262.04) -- (666.55,262.31) .. controls (672.63,262.32) and (677.55,267.26) .. (677.53,273.34) -- (677.45,306.36) .. controls (677.44,312.43) and (672.5,317.35) .. (666.42,317.33) -- (559.16,317.07) .. controls (553.08,317.06) and (548.17,312.12) .. (548.18,306.04) -- cycle ;
\draw  [fill={rgb, 255:red, 255; green, 255; blue, 255 }  ,fill opacity=1 ] (684.26,273.02) .. controls (684.28,266.94) and (689.22,262.03) .. (695.29,262.04) -- (802.55,262.31) .. controls (808.63,262.32) and (813.55,267.26) .. (813.53,273.34) -- (813.45,306.36) .. controls (813.44,312.43) and (808.5,317.35) .. (802.42,317.33) -- (695.16,317.07) .. controls (689.08,317.06) and (684.17,312.12) .. (684.18,306.04) -- cycle ;
\draw  [fill={rgb, 255:red, 255; green, 255; blue, 255 }  ,fill opacity=1 ] (366.26,245.59) -- (376.79,231.62) -- (376.77,240.36) -- (396.02,240.41) -- (396.04,230.16) -- (387.29,230.14) -- (401.32,219.67) -- (415.29,230.21) -- (406.54,230.19) -- (406.52,240.44) -- (425.77,240.48) -- (425.79,231.73) -- (436.26,245.76) -- (425.72,259.73) -- (425.74,250.98) -- (376.74,250.86) -- (376.72,259.61) -- cycle ;
\draw  [fill={rgb, 255:red, 255; green, 255; blue, 255 }  ,fill opacity=1 ] (646.18,245.79) -- (656.71,231.81) -- (656.69,240.56) -- (675.94,240.61) -- (675.97,230.36) -- (667.22,230.34) -- (681.24,219.87) -- (695.22,230.41) -- (686.47,230.38) -- (686.44,240.63) -- (705.69,240.68) -- (705.71,231.93) -- (716.18,245.96) -- (705.64,259.93) -- (705.66,251.18) -- (656.66,251.06) -- (656.64,259.81) -- cycle ;
\draw  [fill={rgb, 255:red, 255; green, 255; blue, 255 }  ,fill opacity=1 ] (333.26,332.02) .. controls (333.28,325.94) and (338.22,321.03) .. (344.29,321.04) -- (451.55,321.31) .. controls (457.63,321.32) and (462.55,326.26) .. (462.53,332.34) -- (462.45,365.36) .. controls (462.44,371.43) and (457.5,376.35) .. (451.42,376.33) -- (344.16,376.07) .. controls (338.08,376.06) and (333.17,371.12) .. (333.18,365.04) -- cycle ;
\draw  [fill={rgb, 255:red, 255; green, 255; blue, 255 }  ,fill opacity=1 ] (615.26,333.02) .. controls (615.28,326.94) and (620.22,322.03) .. (626.29,322.04) -- (733.55,322.31) .. controls (739.63,322.32) and (744.55,327.26) .. (744.53,333.34) -- (744.45,366.36) .. controls (744.44,372.43) and (739.5,377.35) .. (733.42,377.33) -- (626.16,377.07) .. controls (620.08,377.06) and (615.17,372.12) .. (615.18,366.04) -- cycle ;

\draw (527.28,56.38) node  [font=\Large,rotate=-359.67] [align=left] {\begin{minipage}[lt]{37.84pt}\setlength\topsep{0pt}
\begin{center}
Logic
\end{center}

\end{minipage}};
\draw (454.01,180.81) node  [font=\large,rotate=-359.67] [align=left] {Temporal };
\draw (616.01,180.89) node  [font=\large,rotate=-359.67] [align=left] {Strategic };
\draw (328.86,288.69) node  [font=\normalsize,rotate=-0.14] [align=left] {\begin{minipage}[lt]{75.91pt}\setlength\topsep{0pt}
\begin{center}
Linear Temporal\\Logic
\end{center}

\end{minipage}};
\draw (466.86,289.69) node  [font=\normalsize,rotate=-0.14] [align=left] {\begin{minipage}[lt]{84.03pt}\setlength\topsep{0pt}
\begin{center}
Computation Tree\\Logic
\end{center}

\end{minipage}};
\draw (612.86,289.69) node  [font=\normalsize,rotate=-0.14] [align=left] {\begin{minipage}[lt]{77.02pt}\setlength\topsep{0pt}
\begin{center}
Alternating-time \\Temporal Logic
\end{center}

\end{minipage}};
\draw (748.86,291.69) node  [font=\normalsize,rotate=-0.14] [align=left] {\begin{minipage}[lt]{40.72pt}\setlength\topsep{0pt}
\begin{center}
Strategy\\Logic
\end{center}

\end{minipage}};
\draw (397.86,348.69) node  [font=\normalsize,rotate=-0.14] [align=left] {\begin{minipage}[lt]{11.22pt}\setlength\topsep{0pt}
\begin{center}
...
\end{center}

\end{minipage}};
\draw (679.86,349.69) node  [font=\normalsize,rotate=-0.14] [align=left] {\begin{minipage}[lt]{11.22pt}\setlength\topsep{0pt}
\begin{center}
...
\end{center}

\end{minipage}};

\end{tikzpicture}

}
\caption{Logic component insights.}
\label{fig:logic}
\end{figure*}

Given our methodology's intrinsic compositional approach to handling logics, it allows for the addition of further branches in the hierarchy. However, in its initial phase, we selected the two most studied and commonly used branches of logics for verification purposes.

\subsection{Model Checker Interface}


Once the model and logic are chosen, the next step in our methodology is their verification. This is obtained through the \texttt{Model Checker Interface}, presented in~\Cref{fig:model-checker-interface-hierarchy}.
Notice that, each leaf of~\Cref{fig:model-checker-interface-hierarchy} represents a meta-node that can be further decomposed as shown in~\Cref{fig:model-checker-interface} for the case of Strategy Logic.

\tikzset{every picture/.style={line width=0.75pt}} 

\begin{figure*}
\centering
\scalebox{0.35}{
\begin{tikzpicture}[x=0.75pt,y=0.75pt,yscale=-1,xscale=1]

\draw  [fill={rgb, 255:red, 74; green, 144; blue, 226 }  ,fill opacity=0.2 ] (261.5,28.3) -- (817.5,28.3) -- (817.5,390.3) -- (261.5,390.3) -- cycle ;
\draw  [fill={rgb, 255:red, 255; green, 255; blue, 255 }  ,fill opacity=1 ] (437.42,48.12) .. controls (437.37,40.72) and (443.34,34.69) .. (450.74,34.64) -- (607.44,33.74) .. controls (614.84,33.7) and (620.87,39.66) .. (620.91,47.06) -- (621.15,87.26) .. controls (621.19,94.66) and (615.22,100.7) .. (607.82,100.74) -- (451.13,101.64) .. controls (443.72,101.69) and (437.69,95.72) .. (437.65,88.32) -- cycle ;
\draw  [fill={rgb, 255:red, 255; green, 255; blue, 255 }  ,fill opacity=1 ] (496.95,138.39) -- (507.37,124.33) -- (507.42,133.08) -- (526.67,132.96) -- (526.61,122.71) -- (517.86,122.76) -- (531.8,112.18) -- (545.86,122.6) -- (537.11,122.65) -- (537.17,132.9) -- (556.42,132.79) -- (556.37,124.04) -- (566.95,137.98) -- (556.53,152.04) -- (556.48,143.29) -- (507.48,143.57) -- (507.53,152.32) -- cycle ;
\draw  [fill={rgb, 255:red, 255; green, 255; blue, 255 }  ,fill opacity=1 ] (381.13,170.95) .. controls (381.09,163) and (387.5,156.51) .. (395.45,156.47) -- (516.15,155.77) .. controls (524.1,155.72) and (530.58,162.13) .. (530.63,170.09) -- (530.88,213.29) .. controls (530.93,221.24) and (524.52,227.72) .. (516.56,227.77) -- (395.86,228.46) .. controls (387.91,228.51) and (381.43,222.1) .. (381.38,214.15) -- cycle ;
\draw  [fill={rgb, 255:red, 255; green, 255; blue, 255 }  ,fill opacity=1 ] (540.14,171.03) .. controls (540.09,163.08) and (546.5,156.6) .. (554.45,156.55) -- (675.15,155.85) .. controls (683.1,155.81) and (689.59,162.22) .. (689.63,170.17) -- (689.88,213.37) .. controls (689.93,221.32) and (683.52,227.81) .. (675.57,227.85) -- (554.87,228.55) .. controls (546.92,228.59) and (540.43,222.18) .. (540.39,214.23) -- cycle ;
\draw  [fill={rgb, 255:red, 255; green, 255; blue, 255 }  ,fill opacity=1 ] (266.26,283.33) .. controls (266.28,277.26) and (271.22,272.34) .. (277.29,272.35) -- (384.55,272.62) .. controls (390.63,272.63) and (395.55,277.57) .. (395.53,283.65) -- (395.45,316.67) .. controls (395.44,322.74) and (390.5,327.66) .. (384.42,327.65) -- (277.16,327.38) .. controls (271.08,327.37) and (266.17,322.43) .. (266.18,316.35) -- cycle ;
\draw  [fill={rgb, 255:red, 255; green, 255; blue, 255 }  ,fill opacity=1 ] (404.26,284.33) .. controls (404.28,278.26) and (409.22,273.34) .. (415.29,273.35) -- (522.55,273.62) .. controls (528.63,273.63) and (533.55,278.57) .. (533.53,284.65) -- (533.45,317.67) .. controls (533.44,323.74) and (528.5,328.66) .. (522.42,328.65) -- (415.16,328.38) .. controls (409.08,328.37) and (404.17,323.43) .. (404.18,317.35) -- cycle ;
\draw  [fill={rgb, 255:red, 255; green, 255; blue, 255 }  ,fill opacity=1 ] (550.26,284.33) .. controls (550.28,278.26) and (555.22,273.34) .. (561.29,273.35) -- (668.55,273.62) .. controls (674.63,273.63) and (679.55,278.57) .. (679.53,284.65) -- (679.45,317.67) .. controls (679.44,323.74) and (674.5,328.66) .. (668.42,328.65) -- (561.16,328.38) .. controls (555.08,328.37) and (550.17,323.43) .. (550.18,317.35) -- cycle ;
\draw  [fill={rgb, 255:red, 255; green, 255; blue, 255 }  ,fill opacity=1 ] (686.26,284.33) .. controls (686.28,278.26) and (691.22,273.34) .. (697.29,273.35) -- (804.55,273.62) .. controls (810.63,273.63) and (815.55,278.57) .. (815.53,284.65) -- (815.45,317.67) .. controls (815.44,323.74) and (810.5,328.66) .. (804.42,328.65) -- (697.16,328.38) .. controls (691.08,328.37) and (686.17,323.43) .. (686.18,317.35) -- cycle ;
\draw  [fill={rgb, 255:red, 255; green, 255; blue, 255 }  ,fill opacity=1 ] (368.26,256.9) -- (378.79,242.93) -- (378.77,251.68) -- (398.02,251.72) -- (398.04,241.47) -- (389.29,241.45) -- (403.32,230.99) -- (417.29,241.52) -- (408.54,241.5) -- (408.52,251.75) -- (427.77,251.8) -- (427.79,243.05) -- (438.26,257.07) -- (427.72,271.05) -- (427.74,262.3) -- (378.74,262.18) -- (378.72,270.93) -- cycle ;
\draw  [fill={rgb, 255:red, 255; green, 255; blue, 255 }  ,fill opacity=1 ] (648.18,257.1) -- (658.71,243.12) -- (658.69,251.87) -- (677.94,251.92) -- (677.97,241.67) -- (669.22,241.65) -- (683.24,231.18) -- (697.22,241.72) -- (688.47,241.7) -- (688.44,251.95) -- (707.69,251.99) -- (707.71,243.24) -- (718.18,257.27) -- (707.64,271.24) -- (707.66,262.49) -- (658.66,262.37) -- (658.64,271.12) -- cycle ;
\draw  [fill={rgb, 255:red, 255; green, 255; blue, 255 }  ,fill opacity=1 ] (335.26,343.33) .. controls (335.28,337.26) and (340.22,332.34) .. (346.29,332.35) -- (453.55,332.62) .. controls (459.63,332.63) and (464.55,337.57) .. (464.53,343.65) -- (464.45,376.67) .. controls (464.44,382.74) and (459.5,387.66) .. (453.42,387.65) -- (346.16,387.38) .. controls (340.08,387.37) and (335.17,382.43) .. (335.18,376.35) -- cycle ;
\draw  [fill={rgb, 255:red, 255; green, 255; blue, 255 }  ,fill opacity=1 ] (617.26,344.33) .. controls (617.28,338.26) and (622.22,333.34) .. (628.29,333.35) -- (735.55,333.62) .. controls (741.63,333.63) and (746.55,338.57) .. (746.53,344.65) -- (746.45,377.67) .. controls (746.44,383.74) and (741.5,388.66) .. (735.42,388.65) -- (628.16,388.38) .. controls (622.08,388.37) and (617.17,383.43) .. (617.18,377.35) -- cycle ;

\draw (529.28,67.69) node  [font=\Large,rotate=-359.67] [align=left] {\begin{minipage}[lt]{101.5pt}\setlength\topsep{0pt}
\begin{center}
Model Checker\\Interface
\end{center}

\end{minipage}};
\draw (456.01,192.12) node  [font=\large,rotate=-359.67] [align=left] {\begin{minipage}[lt]{88.47pt}\setlength\topsep{0pt}
\begin{center}
Temporal \\Model Checker 
\end{center}

\end{minipage}};
\draw (615.01,192.2) node  [font=\large,rotate=-359.67] [align=left] {\begin{minipage}[lt]{85.07pt}\setlength\topsep{0pt}
\begin{center}
Strategic \\Model Checker
\end{center}

\end{minipage}};
\draw (330.86,300) node  [font=\normalsize,rotate=-0.14] [align=left] {\begin{minipage}[lt]{75.91pt}\setlength\topsep{0pt}
\begin{center}
Linear Temporal\\Logic
\end{center}

\end{minipage}};
\draw (468.86,301) node  [font=\normalsize,rotate=-0.14] [align=left] {\begin{minipage}[lt]{84.03pt}\setlength\topsep{0pt}
\begin{center}
Computation Tree\\Logic
\end{center}

\end{minipage}};
\draw (614.86,301) node  [font=\normalsize,rotate=-0.14] [align=left] {\begin{minipage}[lt]{77.02pt}\setlength\topsep{0pt}
\begin{center}
Alternating-time \\Temporal Logic
\end{center}

\end{minipage}};
\draw (750.86,303) node  [font=\normalsize,rotate=-0.14] [align=left] {\begin{minipage}[lt]{40.72pt}\setlength\topsep{0pt}
\begin{center}
Strategy\\Logic
\end{center}

\end{minipage}};
\draw (399.86,360) node  [font=\normalsize,rotate=-0.14] [align=left] {\begin{minipage}[lt]{11.22pt}\setlength\topsep{0pt}
\begin{center}
...
\end{center}

\end{minipage}};
\draw (681.86,361) node  [font=\normalsize,rotate=-0.14] [align=left] {\begin{minipage}[lt]{11.22pt}\setlength\topsep{0pt}
\begin{center}
...
\end{center}

\end{minipage}};

\end{tikzpicture}

}
\caption{Model Checker Interface insights.}
\label{fig:model-checker-interface-hierarchy}
\end{figure*}
\tikzset{every picture/.style={line width=0.75pt}} 

\begin{figure*}
\centering
\scalebox{0.55}{
\begin{tikzpicture}[x=0.75pt,y=0.75pt,yscale=-1,xscale=1]

\draw  [fill={rgb, 255:red, 74; green, 144; blue, 226 }  ,fill opacity=0.1 ] (349.25,20) -- (831.17,20) -- (831.17,237) -- (349.25,237) -- cycle ;
\draw  [fill={rgb, 255:red, 255; green, 255; blue, 255 }  ,fill opacity=1 ] (501,41.4) .. controls (501,34) and (507,28) .. (514.4,28) -- (671.1,28) .. controls (678.5,28) and (684.5,34) .. (684.5,41.4) -- (684.5,81.6) .. controls (684.5,89) and (678.5,95) .. (671.1,95) -- (514.4,95) .. controls (507,95) and (501,89) .. (501,81.6) -- cycle ;
\draw  [fill={rgb, 255:red, 255; green, 255; blue, 255 }  ,fill opacity=1 ] (559,134) -- (569.5,120) -- (569.5,128.75) -- (588.75,128.75) -- (588.75,118.5) -- (580,118.5) -- (594,108) -- (608,118.5) -- (599.25,118.5) -- (599.25,128.75) -- (618.5,128.75) -- (618.5,120) -- (629,134) -- (618.5,148) -- (618.5,139.25) -- (569.5,139.25) -- (569.5,148) -- cycle ;
\draw  [fill={rgb, 255:red, 255; green, 255; blue, 255 }  ,fill opacity=1 ] (357,165.4) .. controls (357,157.45) and (363.45,151) .. (371.4,151) -- (492.1,151) .. controls (500.05,151) and (506.5,157.45) .. (506.5,165.4) -- (506.5,208.6) .. controls (506.5,216.55) and (500.05,223) .. (492.1,223) -- (371.4,223) .. controls (363.45,223) and (357,216.55) .. (357,208.6) -- cycle ;
\draw  [fill={rgb, 255:red, 255; green, 255; blue, 255 }  ,fill opacity=1 ] (516,166.4) .. controls (516,158.45) and (522.45,152) .. (530.4,152) -- (651.1,152) .. controls (659.05,152) and (665.5,158.45) .. (665.5,166.4) -- (665.5,209.6) .. controls (665.5,217.55) and (659.05,224) .. (651.1,224) -- (530.4,224) .. controls (522.45,224) and (516,217.55) .. (516,209.6) -- cycle ;
\draw  [fill={rgb, 255:red, 255; green, 255; blue, 255 }  ,fill opacity=1 ] (673,167.4) .. controls (673,159.45) and (679.45,153) .. (687.4,153) -- (808.1,153) .. controls (816.05,153) and (822.5,159.45) .. (822.5,167.4) -- (822.5,210.6) .. controls (822.5,218.55) and (816.05,225) .. (808.1,225) -- (687.4,225) .. controls (679.45,225) and (673,218.55) .. (673,210.6) -- cycle ;

\draw (592.75,61.5) node  [font=\Large] [align=left] {\begin{minipage}[lt]{101.5pt}\setlength\topsep{0pt}
\begin{center}
Strategy\\Logic
\end{center}

\end{minipage}};
\draw (431.75,187) node  [font=\Large] [align=left] {Explicit MC};
\draw (747.75,189) node  [font=\Large] [align=left] {Implicit MC};
\draw (590.75,188) node  [font=\Large] [align=left] {Abstract MC};

\end{tikzpicture}

}
\caption{Meta-node for Strategy Logic.}
\label{fig:model-checker-interface}
\end{figure*}

To better comprehend the role of such a component, it is noteworthy that our methodology incorporates a selection mechanism connected to the model checker interface. This mechanism enables the discernment of the appropriate model checker usage, considering the selected model and logic. The interface is configured to analyse the model description and logical formula, determining the class of model among a set of predefined model classes to which the model belongs (\texttt{Model} component) and the class of logic among a set of predefined logic classes to which the formula belongs (\texttt{Logic} component). Then, given the model and logic, the model checker interface selects the most efficient verification method. For instance, by assuming the number of states as main parameter of the problem, the model checker interface could select an explicit method for models with less than fifty states, an implicit method for models with less than one-hundred states, and an abstract method with more than one-hundred states.




\input{figures/flow}

\subsection{User interface}


In this section, we focus on the way the end-user can interact with \vitamin. 
In our approach, we categorise end-users based on their expertise in formal methods. Specifically, we distinguish between \texttt{Expert} and \texttt{Non-Expert} users.
In both cases, being end-users of the system, they would benefit from a Graphical User Interface (GUI) to guide them through the verification of their systems.

In~\Cref{fig:flowchart}, we present an ideal interaction flow for both types of users. 

On the left, we depict the interaction involving an expert user, which is more straightforward and direct. Since the user is experienced with formal methods, they can simply upload the corresponding files for the model and formula through a GUI. These files must be compatible with the format expected by the specific instantiation. Once these files are provided as input, the verification continues.

On the right, the interaction is guided, as the non-expert user lacks experience with formal methods. To address this, our methodology solicits details and information about the model and the formula. These information can be gathered through a sequence of questions to the user. For instance, the user may be asked about the number of agents he/she thinks to employ in the MAS, or, how many and which kind of actions are available for such agents. These examples serve only as illustrations, demonstrating the system's potential to guide non-expert users.  By following such a constrained step-by-step process, a non-expert user can interact in a more natural way. Once information about the model is supplied, we may continue by seeking additional details about the formula. 
The communication in this phase can also be handled through natural language. Questions during this step may inquire about the specific property of interest, which might involve considering temporal information. It is worth noting that this step is inspired by what is commonly done in the FRET framework~\cite{DBLP:conf/refsq/GiannakopoulouP20a} (and similar ones), where users can describe formal properties in a constrained natural language, and it is the system's responsibility to generate the corresponding formal property in the chosen formalism.

At the conclusion of the guided process, akin to the expert user, the non-expert user can proceed with the verification steps illustrated in~\Cref{fig:flowchart}. 
This process initiates with a parsing step, wherein the model and formula are parsed, leading to the creation of an internal representation.
Following the parsing step, the verification process advances with the careful selection of an appropriate model checker to address the verification of the provided model against the given formula. Subsequently, the verification result is returned to the user.

\section{Implementation}


In this section, we focus on the instantiation of \vitamin and what the tool currently supports.

\vitamin is implemented in Python and its Graphical User Interface is accessible through a web browser (\url{https://vitamin-app.streamlit.app/}), which makes the tool cross-platform. This accessibility is facilitated by utilising the Streamlit\footnote{\url{https://streamlit.io/}} Python library, which supports the transparent sharing of Python programs via HTTP protocol.

Furthermore, the source code of \vitamin can be found at \url{https://github.com/VadimMalvone/VITAMIN}. 
Nonetheless, it is worth mentioning that the actual implementation of the modules is beyond the scope of this paper, as its goal is to present the engineering and architecture of the \vitamin framework.

As mentioned in the paper, \vitamin supports the interaction with both expert and non-expert users.

\subsection{Non-Expert user experience in \vitamin}

\begin{figure}[ht!]
    \centering
    \includegraphics[width=0.7\linewidth]{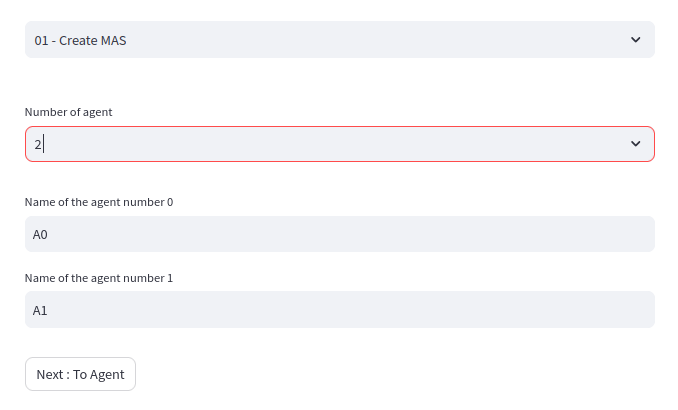}
    \caption{\vitamin asks the user for the agents to employ in the model of the MAS.}
    \label{fig:number_agents}
\end{figure}

Figure~\ref{fig:number_agents} reports a screenshot of \vitamin's GUI where the user is asked about the number of agents to employ in the MAS to verify. In this specific example, the user wants to create a MAS comprising two agents, named \texttt{A0} and \texttt{A1}. 

\begin{figure}[ht!]
    \centering
    \includegraphics[width=0.7\linewidth]{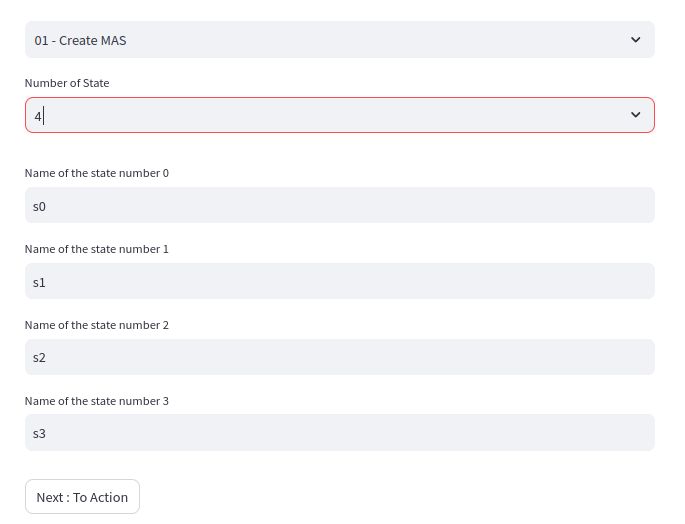}
    \caption{\vitamin asks the user for the states of the model of the MAS.}
    \label{fig:number_states}
\end{figure}

After the number of agents have been given to the system, the process goes on with the number of states. As reported in Figure~\ref{fig:number_states}, the user inserts the number of states he/she wishes to add in the model of the MAS under analysis. In this specific case, the user chooses to add four states, which are then named: \texttt{S0}, \texttt{S1}, \texttt{S2}, and \texttt{S3}.

\begin{figure}[ht!]
    \centering
    \includegraphics[width=0.7\linewidth]{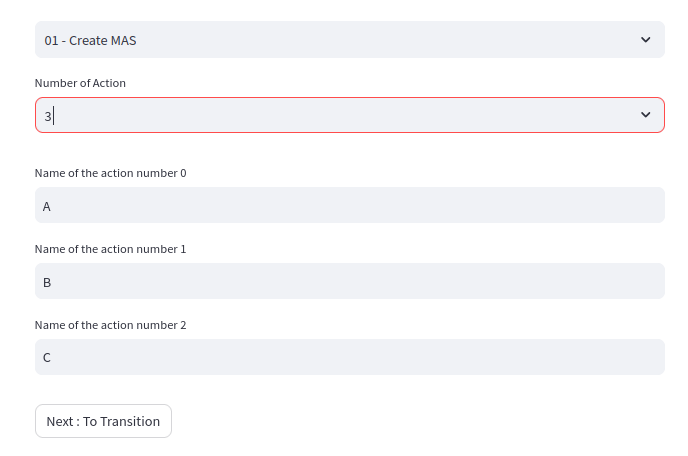}
    \caption{\vitamin asks the user for the actions the agents can perform.}
    \label{fig:number_actions}
\end{figure}

Once both agents and states have been gathered, \vitamin asks for the actions that the agents can perform in the states. Figure~\ref{fig:number_actions} reports the step where \vitamin asks the user for the actions to be assigned to the previously created agents, in the previously added states. In this specific scenario, the user decides that the agents can perform three actions, that are named: \texttt{A}, \texttt{B}, and \texttt{C}.

\begin{figure}[ht!]
    \centering
    \includegraphics[width=0.6\linewidth]{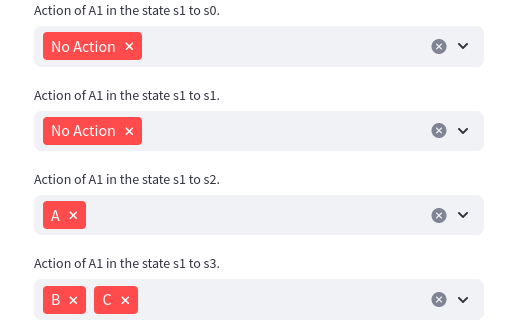}
    \caption{\vitamin asks the user for transitions amongst states.}
    \label{fig:transitions}
\end{figure}

\begin{figure}[ht!]
    \centering
    \includegraphics[width=0.7\linewidth]{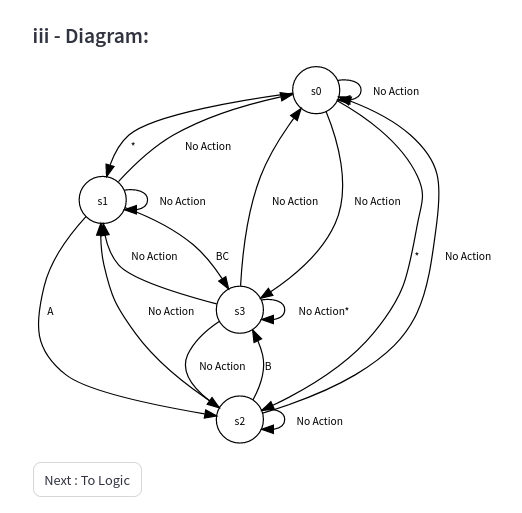}
    \caption{\vitamin shows the graph resulting from the information gathered in the previous steps to allow the user to validate it.}
    \label{fig:graph}
\end{figure}

Now that the agents have actions to perform, \vitamin requires the user to specify which actions can be performed by which agent in which state. This process allows a natural definition of transitions amongst states through agents' actions. The transitions are reported by the user through \vitamin's GUI. In~\Cref{fig:transitions}, we report only a subset of \texttt{A1}'s transitions to improve readability. In this specific scenario, agent \texttt{A1} can perform only action \texttt{A} in state \texttt{S1} to move to state \texttt{S2}, and actions \texttt{B} and \texttt{C} to move to state \texttt{S3}. 
For the remaining transitions, both for \texttt{A0} and \texttt{A1}, the reader can refer to~\Cref{fig:graph}.

At this point, \vitamin has all the information it needs to represent the graph of the model created by the user. To allow the user to validate the resulting model, \vitamin shows the graph result to the user, as reported in Figure~\ref{fig:graph}. \vitamin supports the graphical visualisation of the model and allows the user to validate it before moving on in the verification process. 

\begin{figure}[ht!]
    \centering
    \includegraphics[width=0.7\linewidth]{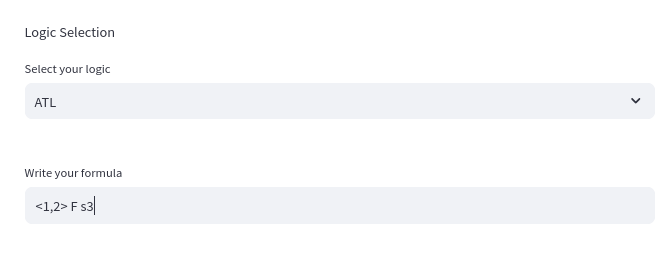}
    \caption{\vitamin asks the user for the formula to verify on the previously constructed model of the MAS.}
    \label{fig:formula}
\end{figure}

Finally, given the model produced step-by-step by guiding the user through all the details needed to populate the model, \vitamin can conclude the process by asking the formula to verify on the obtained model. Figure~\ref{fig:formula} reports the last step of the \vitamin's process where the user can specify the formal property to verify on the model. In this specific case, the user decides to specify an ATL formula and, in particular, to verify whether the agents can reach state \texttt{S3} by collaborating. Such a property is verified then by \vitamin and concluded as satisfied on the current model of the MAS.

\vitamin's current instantiation requires the formula to be specified, however, a step-by-step mechanism could be employed as well. Nonetheless, differently from the model scenario, the formula may require additional engineering since it may largely depend on the formalism chosen by the user.



\subsection{Expert user experience in \vitamin}


In this section, we show how an expert user interacts with \vitamin. Specifically, this is achieved by the definition of the model of the MAS and the formal property the user wishes to verify. Differently from the non-expert user, \vitamin does not guide the expert user, but it expects the model and formula as input.

\begin{figure}[ht!]
    \centering
    \includegraphics[width=0.7\linewidth]{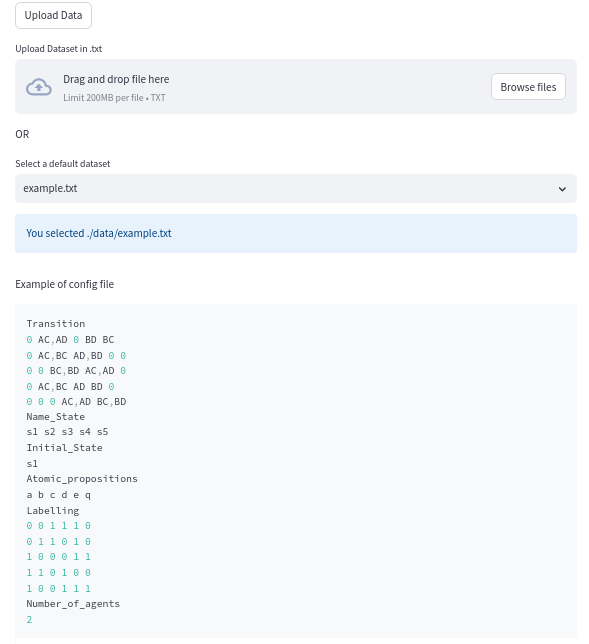}
    \caption{Screenshot of \vitamin's GUI with an example.}
    \label{fig:model-example}
\end{figure}

As discussed before, \vitamin is born to handle different kind of models and logics. Thanks to its design, it is not limited to any specific model (resp., logic) since each model (resp., logic) can be seen as an independent component of the system. However, to make an example, we show how to define a model of a MAS as a CGS in \vitamin.
Figure~\ref{fig:model-example} reports a screenshot of \vitamin's GUI where the user can upload the CGS of the MAS to verify. The model has to follow a specific format that has been chosen for the definition of CGSs in \vitamin. Naturally, because of its compositionality, these format choices related exclusively to how \vitamin handles CGSs and does not concern the development of other formalisms for the representation of models. That is, if \vitamin supported Interpreted Systems (like MCMAS), it would be free to choose the format that best suit such models, without be concerned on how CGSs are modelled, and \textit{vice versa}.

\begin{figure}[ht!]
    \centering
    \includegraphics[width=0.7\linewidth]{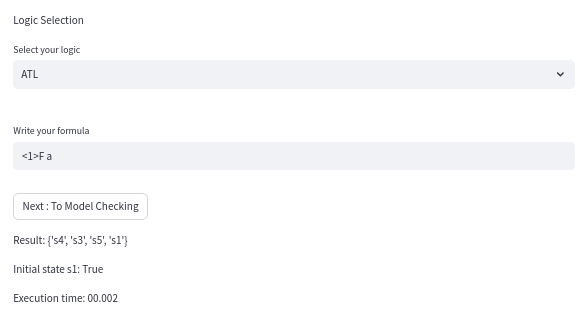}
    \caption{Screenshot of \vitamin's GUI with an example of property to be verified.}
    \label{fig:formula-example}
\end{figure}

After the model has been uploaded by the user, \vitamin expects the formula to be verified on the latter. This step, similarly to the non-expert user scenario, is obtained by letting the user fill a field box in the \vitamin's GUI. Figure~\ref{fig:formula-example} reports a screenshot of the GUI where the user fills the box with the formula of interest to verify on the model.

Once both the model and formula are given, the verification process may start and the result of the verification is returned back to the user.

\section{Conclusions and Future Work}

In this paper, we introduced \vitamin, a comprehensive and versatile framework designed for model checking of Multi-Agent Systems and beyond. Our emphasis was on the engineering aspects and decisions made during the development of \vitamin.

We acknowledge that \vitamin is an ongoing project that will necessitate additional refinement, but we recognise that its current state already represents a noteworthy advancement in the realm of tools for the formal verification of MAS. This is especially notable given its potential for further study and extension, facilitated by its inherent compositionality.

It is important to note that \vitamin is currently in a prototype stage. Certain aspects presented in this paper
such as Natural Language Processing (NLP) support for non-expert users and the \texttt{Model Checker Interface} part, are still in development. However, everything related to verification and compositional representation in \vitamin has already been implemented and tested across various scenarios, each highlighting different models and formulas for verification.

Our future plans include the continued expansion of \vitamin, along with sharing it with the MAS community. Additionally, we intend to present its extensions in future research endeavours, exploring possible instantiations of models and logics within the tool. While this work has primarily focused on \vitamin's engineering and architecture, future research will delve into specific instantiations and applications (of what we called the \vitamin's components).

\subsubsection*{Acknowledgements.} We thank Hi! Paris (\url{www.hi-paris.fr}) for their support in the initial development of \vitamin. We also thank all the students that collaborated on the project by developing some of its components. Finally, we thank Aniello Murano for his valuable feedback and support to this project. 


%
%
%
\bibliographystyle{splncs04}
\bibliography{main}

\begin{thebibliography}{10}
\providecommand{\url}[1]{\texttt{#1}}
\providecommand{\urlprefix}{URL }
\providecommand{\doi}[1]{https://doi.org/#1}

\bibitem{DBLP:journals/jacm/AlurHK02}
Alur, R., Henzinger, T.A., Kupferman, O.: Alternating-time temporal logic. J. {ACM}  \textbf{49}(5),  672--713 (2002). \doi{10.1145/585265.585270}, \url{https://doi.org/10.1145/585265.585270}

\bibitem{BelardinelliFM23}
Belardinelli, F., Ferrando, A., Malvone, V.: An abstraction-refinement framework for verifying strategic properties in multi-agent systems with imperfect information. Artif. Intell.  \textbf{316},  103847 (2023). \doi{10.1016/j.artint.2022.103847}, \url{https://doi.org/10.1016/j.artint.2022.103847}

\bibitem{DBLP:conf/ijcai/BelardinelliJKM19}
Belardinelli, F., Jamroga, W., Kurpiewski, D., Malvone, V., Murano, A.: Strategy logic with simple goals: Tractable reasoning about strategies. In: Kraus, S. (ed.) Proceedings of the Twenty-Eighth International Joint Conference on Artificial Intelligence, {IJCAI} 2019, Macao, China, August 10-16, 2019. pp. 88--94. ijcai.org (2019). \doi{10.24963/IJCAI.2019/13}, \url{https://doi.org/10.24963/ijcai.2019/13}

\bibitem{BelardinelliLMY22}
Belardinelli, F., Lomuscio, A., Malvone, V., Yu, E.: Approximating perfect recall when model checking strategic abilities: Theory and applications. J. Artif. Intell. Res.  \textbf{73},  897--932 (2022). \doi{10.1613/jair.1.12539}, \url{https://doi.org/10.1613/jair.1.12539}

\bibitem{DBLP:books/cu/Chellas80}
Chellas, B.F.: Modal Logic - An Introduction. Cambridge University Press (1980). \doi{10.1017/CBO9780511621192}, \url{https://doi.org/10.1017/CBO9780511621192}

\bibitem{DBLP:conf/lop/ClarkeE81}
Clarke, E.M., Emerson, E.A.: Design and synthesis of synchronization skeletons using branching-time temporal logic. In: Kozen, D. (ed.) Logics of Programs, Workshop, Yorktown Heights, New York, USA, May 1981. Lecture Notes in Computer Science, vol.~131, pp. 52--71. Springer (1981). \doi{10.1007/BFB0025774}, \url{https://doi.org/10.1007/BFb0025774}

\bibitem{DBLP:journals/ase/DennisFWB12}
Dennis, L.A., Fisher, M., Webster, M.P., Bordini, R.H.: Model checking agent programming languages. Autom. Softw. Eng.  \textbf{19}(1),  5--63 (2012). \doi{10.1007/S10515-011-0088-X}, \url{https://doi.org/10.1007/s10515-011-0088-x}

\bibitem{DBLP:journals/corr/abs-1102-4225}
Dima, C., Tiplea, F.L.: Model-checking {ATL} under imperfect information and perfect recall semantics is undecidable. CoRR  \textbf{abs/1102.4225} (2011), \url{http://arxiv.org/abs/1102.4225}

\bibitem{DBLP:books/mit/FHMV1995}
Fagin, R., Halpern, J.Y., Moses, Y., Vardi, M.Y.: Reasoning About Knowledge. {MIT} Press (1995). \doi{10.7551/MITPRESS/5803.001.0001}, \url{https://doi.org/10.7551/mitpress/5803.001.0001}

\bibitem{FerrandoM22}
Ferrando, A., Malvone, V.: Towards the combination of model checking and runtime verification on multi-agent systems. In: Dignum, F., Mathieu, P., Corchado, J.M., de~la Prieta, F. (eds.) Advances in Practical Applications of Agents, Multi-Agent Systems, and Complex Systems Simulation. The {PAAMS} Collection - 20th International Conference, {PAAMS} 2022, L'Aquila, Italy, July 13-15, 2022, Proceedings. Lecture Notes in Computer Science, vol. 13616, pp. 140--152. Springer (2022). \doi{10.1007/978-3-031-18192-4\_12}, \url{https://doi.org/10.1007/978-3-031-18192-4\_12}

\bibitem{FerrandoM23}
Ferrando, A., Malvone, V.: Towards the verification of strategic properties in multi-agent systems with imperfect information. In: Agmon, N., An, B., Ricci, A., Yeoh, W. (eds.) Proceedings of the 2023 International Conference on Autonomous Agents and Multiagent Systems, {AAMAS} 2023, London, United Kingdom, 29 May 2023 - 2 June 2023. pp. 793--801. {ACM} (2023). \doi{10.5555/3545946.3598713}, \url{https://dl.acm.org/doi/10.5555/3545946.3598713}

\bibitem{DBLP:conf/refsq/GiannakopoulouP20a}
Giannakopoulou, D., Pressburger, T., Mavridou, A., Rhein, J., Schumann, J., Shi, N.: Formal requirements elicitation with {FRET}. In: Sabetzadeh, M., Vogelsang, A., Abualhaija, S., Borg, M., Dalpiaz, F., Daneva, M., Condori{-}Fern{\'{a}}ndez, N., Franch, X., Fucci, D., Gervasi, V., Groen, E.C., Guizzardi, R.S.S., Herrmann, A., Horkoff, J., Mich, L., Perini, A., Susi, A. (eds.) Joint Proceedings of {REFSQ-2020} Workshops, Doctoral Symposium, Live Studies Track, and Poster Track co-located with the 26th International Conference on Requirements Engineering: Foundation for Software Quality {(REFSQ} 2020), Pisa, Italy, March 24, 2020. {CEUR} Workshop Proceedings, vol.~2584. CEUR-WS.org (2020), \url{https://ceur-ws.org/Vol-2584/PT-paper4.pdf}

\bibitem{DBLP:journals/cacm/Keller76}
Keller, R.M.: Formal verification of parallel programs. Commun. {ACM}  \textbf{19}(7),  371--384 (1976). \doi{10.1145/360248.360251}, \url{https://doi.org/10.1145/360248.360251}

\bibitem{DBLP:conf/atal/KurpiewskiJK19}
Kurpiewski, D., Jamroga, W., Knapik, M.: {STV:} model checking for strategies under imperfect information. In: Elkind, E., Veloso, M., Agmon, N., Taylor, M.E. (eds.) Proceedings of the 18th International Conference on Autonomous Agents and MultiAgent Systems, {AAMAS} '19, Montreal, QC, Canada, May 13-17, 2019. pp. 2372--2374. International Foundation for Autonomous Agents and Multiagent Systems (2019), \url{http://dl.acm.org/citation.cfm?id=3332116}

\bibitem{DBLP:journals/sttt/LomuscioQR17}
Lomuscio, A., Qu, H., Raimondi, F.: {MCMAS:} an open-source model checker for the verification of multi-agent systems. Int. J. Softw. Tools Technol. Transf.  \textbf{19}(1),  9--30 (2017). \doi{10.1007/S10009-015-0378-X}, \url{https://doi.org/10.1007/s10009-015-0378-x}

\bibitem{DBLP:journals/tocl/MogaveroMPV14}
Mogavero, F., Murano, A., Perelli, G., Vardi, M.Y.: Reasoning about strategies: On the model-checking problem. {ACM} Trans. Comput. Log.  \textbf{15}(4),  34:1--34:47 (2014). \doi{10.1145/2631917}, \url{https://doi.org/10.1145/2631917}

\bibitem{DBLP:conf/aose/NguyenPBPT09}
Nguyen, C.D., Perini, A., Bernon, C., Pav{\'{o}}n, J., Thangarajah, J.: Testing in multi-agent systems. In: Gleizes, M., G{\'{o}}mez{-}Sanz, J.J. (eds.) Agent-Oriented Software Engineering {X} - 10th International Workshop, {AOSE} 2009, Budapest, Hungary, May 11-12, 2009, Revised Selected Papers. Lecture Notes in Computer Science, vol.~6038, pp. 180--190. Springer (2009). \doi{10.1007/978-3-642-19208-1\_13}, \url{https://doi.org/10.1007/978-3-642-19208-1\_13}

\bibitem{DBLP:conf/focs/Pnueli77}
Pnueli, A.: The temporal logic of programs. In: 18th Annual Symposium on Foundations of Computer Science, Providence, Rhode Island, USA, 31 October - 1 November 1977. pp. 46--57. {IEEE} Computer Society (1977). \doi{10.1109/SFCS.1977.32}, \url{https://doi.org/10.1109/SFCS.1977.32}

\bibitem{DBLP:journals/jcss/Reif84}
Reif, J.H.: The complexity of two-player games of incomplete information. J. Comput. Syst. Sci.  \textbf{29}(2),  274--301 (1984). \doi{10.1016/0022-0000(84)90034-5}, \url{https://doi.org/10.1016/0022-0000(84)90034-5}

\bibitem{DBLP:conf/atal/Winikoff17}
Winikoff, M.: Debugging agent programs with why?: Questions. In: Larson, K., Winikoff, M., Das, S., Durfee, E.H. (eds.) Proceedings of the 16th Conference on Autonomous Agents and MultiAgent Systems, {AAMAS} 2017, S{\~{a}}o Paulo, Brazil, May 8-12, 2017. pp. 251--259. {ACM} (2017), \url{http://dl.acm.org/citation.cfm?id=3091166}

\end{thebibliography}

\end{document}